\documentclass[traditabstract]{aa} 
\usepackage{graphicx}
\usepackage{txfonts}
\usepackage{longtable}
\usepackage{lscape}
\usepackage{rotating}
\usepackage{subfigure}
\usepackage{natbib}

\begin{document}
   \title{Lithium depletion in solar-like stars: no planet connection}


   \author{Patrick Baumann\inst{1}, Iv\'an Ram\'irez\inst{1}, Jorge Mel\'endez\inst{2}, 
   	   Martin Asplund\inst{1}, and Karin Lind\inst{3}
          }

   \institute{Max Planck Institute for Astrophysics, Postfach 1317, 85741
   Garching, Germany\\
              \email{pbaumann, ivan, asplund@mpa-garching.mpg.de}
         \and
             Centro de Astrofisica da Universidade do Porto, Rua das Estrelas,
	     4150-762 Porto, Portugal\\
             \email{jorge@astro.up.pt}
	 \and
	     European Southern Observatory(ESO), Karl-Schwarzschild-Str. 2, 85748 Garching, Germany\\
	     \email{klind@eso.org}
             }

   \date{Received \dots; accepted \dots}

 
  \abstract
   {
We have determined precise stellar parameters and lithium abundances in a sample of 117 stars with basic properties very
similar to the Sun. This sample selection reduces biasing effects and systematic errors in the analysis.
We estimate the ages of our sample stars mainly from isochrone fitting but also from measurements of rotation period and X-ray luminosity and test the connection between lithium abundance, age, and stellar parameters. We find strong evidence for increasing lithium depletion with age. Our sample includes 14 stars that are known to host planets and it does not support recent claims that planet-host stars have experienced more lithium depletion than stars without planets. We find the solar lithium abundance normal for a star of its age, mass, and metallicity. Furthermore, we analyze published data for 82 stars that were reported to support an enhanced lithium depletion in planet hosts. We show that those stars in fact follow an age trend very similar to that found with our sample and that the presence of giant planets is not related to low lithium abundances. Finally, we discuss the systematic biases that led to the incorrect conclusion of an enhanced lithium depletion in planet-host stars.

   }
   {}


   \authorrunning{P. Baumann et al.}
   \titlerunning{Lithium depletion in solar-like stars}

   \maketitle
%

\section{Introduction}

The lithium abundances of solar-like stars in the solar neighborhood spread over more than
two orders of magnitude, which is much larger than the range of abundances seen for other elements \cite[e.g.,][]{Reddy-03}. The Sun, in particular, has a very low lithium abundance compared to many nearby solar analogs \cite[e.g.,][]{Lambert-Reddy-04}. Furthermore, the photospheric solar lithium abundance is about 160 times lower than that measured in meteorites \cite[$\log\epsilon_{\rm Li,\odot}=1.05\pm0.10$ dex\footnote{We use the standard notation $\log\epsilon_{\rm X}=\log\frac{n_{\rm X}}{n_{\rm H}}+12$, where $n_{\rm X}$ and $n_{\rm H}$ are the the number densities of element X and hydrogen, respectively.\\ Also, for metallicities we use the common abbreviation [Fe/H]$=\log\epsilon_{\rm Fe}-\log\epsilon_{\rm Fe}^{\odot}$.}, $\log\epsilon_{\rm Li,met}=3.26\pm0.05$ dex; both values are from][]{Asplund-09}. This difference between the current solar and protosolar values is not predicted by standard stellar evolution models \cite[e.g.,][]{Dantona-Mazzitelli-84}.

The wide range of observed lithium abundances in nearby solar-like stars is most likely due to a dependency between $\log\epsilon_{\rm Li}$ and the star's age and mass \cite[e.g.,][]{Montalban-Schatzman-00, Charbonnel-Talon-05, Xiong-Deng-09, DoNascimento-09}. Lithium is easily destroyed by proton capture reactions in stellar interiors. Thus, if lithium is transported between the chemically mixed outer convection zone and deeper lying regions with temperatures that are high enough for lithium destruction, the photospheric abundance will decrease with time. Diffusion probably contributes to the lowering of the surface lithium abundance throughout the main-sequence stage.
This would explain why the photospheric solar abundance is much smaller than the meteoritic one.
We expect an enhanced lithium depletion in stars with larger convection zones on the main sequence as well as in stars with a higher degree of differential rotation between the radiative core and the convective envelope (see below). The reason is that lithium is only depleted as it moves to deeper and therefore hotter regions of a star, where the temperature is high enough (about 2.5 million K) for proton capture \cite[see, e.g.,][]{Pinsonneault-97}.

Recently, it has been suggested that the presence of planets around a star could affect the evolution of the photospheric lithium abundance \cite[e.g.,][]{Bouvier-08}. A long-lasting star-disk interaction during the star's pre-main sequence phase could slow down the host-star's rotation and therefore increase the degree of differential rotation between the star's core and envelope. Rotationally-driven mixing is then enhanced, thus destroying more lithium than in stars without planets because fast rotators evolve with little core-envelope decoupling. Planet migration affects the star's angular momentum, which might also have an impact on $\log\epsilon_{\rm Li}$. Finally, the ingestion of a planet can increase the surface lithium abundance \cite[e.g.,][]{Montalban-Rebolo-02, Israelian-01}.

The possibility of a lithium-planet connection is subject of ongoing discussions. Recent work by \cite{Gonzalez-08}, \cite{Gonzalez-Carlson-Tobin-10}, \cite{Castro-08}, and \cite{Israelian-09} suggests a possible $\log\epsilon_{\rm Li}$-planet dependency, whereas \cite{Ryan-00} and \cite{Luck-Heiter-06} find that stars with planets show the same lithium distribution as the comparison field stars. \cite{Takeda-07,Takeda-10} describe the stellar angular momentum as the crucial factor that determines the lithium abundance of solar-like stars and find that slow rotators
show an enhanced lithium depletion. Planets \emph{could} be the reason for a slow rotation, but they were not able to draw firm conclusions due to the low number of planet hosts in their sample and the fact that their use of the star's projected rotational velocity, $v \sin i$, instead of measured rotation periods introduces additional uncertainty, since the inclination angle $i$ is unknown.

From a practical point of view, an enhanced lithium depletion in planet-hosts would be greatly beneficial for the search for extrasolar planets, because all known methods for extrasolar planet detection (e.g., radial velocity, transits, or microlensing) are very time consuming. With an enhanced lithium depletion, however, one could preselect planet-host candidates just by obtaining the stars' chemical composition.\\

In this paper, we derive lithium abundances and ages for a sample of solar-type stars to examine whether there is a correlation between lithium and age as well as a possible connection between lithium and planets. We also examine lithium abundances and ages for the solar analog sample of \cite{Israelian-09}, who claim to have found evidence for an enhanced lithium depletion in planet-host stars.


\section{Observations \& analysis}\label{chap:observations}

Our sample consists of 117 solar-like stars selected from the Hipparcos catalog
\cite[][]{Perryman-97} as in \cite{Melendez-Ramirez-07}.
They where observed at the McDonald (Robert G. Tull coud\'e spectrograph on the 2.7m
Harlan J. Smith telescope; RGT), Las Campanas (MIKE spectrograph on the 6.5m Magellan Clay
telescope), and La Silla (HARPS spectrograph on the 3.6m ESO telescope) observatories.
Our few solar twins observed at Keck \cite[][]{Melendez-Dodds-Eden-Robles-06} are not discussed here since they are already included in the McDonald sample.

The RGT and MIKE data (spectra as well as stellar parameters) are from \citet[][hereafter R09]{Ramirez-Melendez-Asplund-09} and \citet[][hereafter M09]{Melendez-09, Melendez-10}, respectively; two stars (HIP10215 and HIP79672) are part of both samples. HARPS spectra for 12 more stars were obtained from the ESO archive, while spectra for 6 other stars were obtained from the S$^4$N database \cite[][]{AllendePrieto-04}\footnote{The Spectroscopic Survey of Stars in the Solar Neighbourhood (S$^4$N) data and more detailed information can be found at http://hebe.as.utexas.edu/s4n/}. One of the objects (HIP80337) occurs in both the HARPS and the S$^4$N samples (we use the HARPS parameters, because they have the smaller uncertainties), so that we have 17 additional stars. The spectra for these stars were analyzed in an identical fashion as in R09 (see below). All sub-samples contain one or more solar reference objects for normalization: R09 used the light reflected from the asteroids Ceres and Vesta, M09 used Vesta, and for the stars added in this work, spectra of asteroid Ceres, Jupiter's moon Ganymedes, and the Moon were used. Table \ref{table:obs} shows the specifications of each sub-sample, where the last two lines describe the additional data from this work. All spectra have a signal-to-noise ratio (S/N) greater than about 200, which allows the determination of lithium abundances as low as solar. Note that three stars in our sample are also included in M09 (HIP79672) and R09 (HIP14614 and HIP42438). For the further analysis, we use the  parameters with the smaller uncertainties.
\begin{table}[!ht]
\centering                   
\caption{Specifications for the different sub-samples.}
\begin{tabular}{c c c c c c}     
\hline\hline                 
sample & instrument/ & wavelength          & spectral                   & number\\   
 name  &  telescope  & coverage            & resolution                 &  of   \\
       &             & (in $\AA$) & $R=\Delta \lambda/\lambda$ & stars \\
\hline                       
  R09  & RGT /  McDonald & 3800-9125  & 60,000 & 63\\   
  M09  & MIKE / Magellan & 3400-10000 & 65,000 & 42\\
  this & RGT / McDonald, & 3800-9125  & 45,000--& 18\\
  work & HARPS / ESO     & 3783-6865  & 110,000 &  \\
\hline                                  
\end{tabular}
\label{table:obs}
\end{table}

The HARPS and S$^4$N data were analyzed by first measuring \ion{Fe}{i} and \ion{Fe}{ii} equivalent widths (EWs), where a line list of 45 iron lines (34 \ion{Fe}{i} and 11 \ion{Fe}{ii} lines) within the wavelength range from 4445 to 8294 $\AA$ was used. The Fe lines have a broad range of excitation potentials (from $\sim 0.1$ to $\sim 4.6$ eV). The line list adopted is from Scott et al. \cite[in preparation, see also][]{Asplund-09}. To calculate the iron abundances ([Fe/H]) from the \ion{Fe}{i}/\ion{Fe}{ii} lines, we used the spectrum synthesis program \emph{MOOG} \cite[][]{Sneden-73} and ATLAS 9 model atmospheres without convective overshoot \cite[e.g.,][]{Kurucz-93}. We iteratively assigned the stellar parameters effective temperature, surface gravity, and
microturbulent velocity by forcing simultaneously \ion{Fe}{i} excitation equilibrium and \ion{Fe}{i}/\ion{Fe}{ii} ionization equilibrium. For a more detailed description see \cite{Ramirez-Melendez-Asplund-09}.
With the method described above, we derived the following mean errors: $\sigma(\mathrm{T}_{\mathrm{eff}})=40\,$K, $\sigma(\log g )=0.06$ dex, and $\sigma([\mathrm{Fe/H}])=0.025$ dex.

Stellar ages and masses were determined primarily from the location of stars on the
theoretical HR-diagram ($T_\mathrm{eff}$ vs.\ $\log g$) as compared to theoretical
predictions based on stellar evolution calculations (isochrones). We produced a grid
of Y$^2$ isochrones \cite[e.g.,][]{Yi-01} with steps of 0.01\,dex in
metallicity around the solar value. For each star, we computed the age probability
distribution of all isochrone points included within a 3-$\sigma$ radius from the
observed stellar parameters. The adopted mean age and 1-$\sigma$ Gaussian-like upper
and lower limits were derived from the age probability distribution \cite[e.g.,][]{Lachaume-99, Reddy-03}. Although isochrone ages of unevolved stars are typically unreliable, the high precision of our stellar parameters allowed us to infer reasonably accurate absolute isochrone ages, even for stars as young as $\sim3$\,Gyr; relative ages are naturally even better determined. For most stars younger than about $3\,$Gyr only upper limits could be determined. For these younger stars, we adopted ages based on measurements of rotational periods \cite[][]{Gaidos-Henry-Henry-00, Barnes-07} and X-ray luminosity \cite[][]{Agueros-09} along with the rotation-age relation by \cite{Barnes-07} and the X-ray luminosity-age relation by \cite{Guinan-Engle-09}. In a few cases of stars in the intermediate age region ($2-4\,$Gyr), an average of isochrone and rotational ages was determined to improve the age estimate. For stars with very unreliable isochrone ages we used the activity-based ages by \cite{Mamajek-Hillenbrand-08} and \cite{Saffe-Gomez-Chavero-05}. Our adopted ages and the methods to obtain them are listed in Table \ref{table:parameters}.

\begin{table*}[ht]
\begin{center}
\caption{Age and lithium abundance of solar twins in open clusters of near solar metallicity. Data are from the compilation by \cite{Sestito-Randich-05}.}
\begin{tabular}{c c c c c c c c c c c c c c c c c}     
\hline\hline                 
Cluster & Age in Gyr & $\log\epsilon_{\rm Li}$ & $\sigma(\log\epsilon_{\rm Li})$ & [Fe/H] & Source \\
\hline
IC2602 \& IC2391& 0.030 & 2.9  & 0.1  & -0.05 & \cite{Randich-01}     \\		   
Pleiades        & 0.07  & 2.8  & 0.1  & -0.03 & \cite{Soderblom-93}   \\  
Blanco 1        & 0.10  & 2.9  & 0.1  & +0.04 & \cite{Ford-Jeffries-Smalley-05}\\
M34 (NGC1039)   & 0.25  & 2.8  & 0.1  & +0.07 & \cite{Jones-97}	    \\
Coma Berenices  & 0.6   & 2.4  & 0.15 & -0.05 & \cite{Ford-01}	    \\		   
NGC762          & 2.0   & 2.1  & 0.1  & +0.01 & \cite{Sestito-Randich-Pallavicini-04}\\		   
M67             & 3.9   & 1.2  & 0.5  & +0.05 & \cite{Pasquini-08}    \\				   
\hline
\end{tabular}
\label{table:clusters}
\end{center}
\end{table*}

Using our stellar parameters as well as those in R09 and M09, we derived the lithium
abundances for all 117 stars with line synthesis using \emph{MOOG}. For this
purpose we generated a line list from 6697 to 6717 $\AA$, i.e. 10 $\AA$ around the lithium doublet at 6707.8 $\AA$. The whole wavelength range was synthesized with \emph{MOOG}, where hyperfine splitting and the Li-doublet were taken into account. Knowing the basic stellar parameters, we were able to fit the lithium line including the effects of the projected rotational velocity $v \sin i$ and the microturbulent and macroturbulent velocities. We derived lithium abundances with a mean error of $\sigma=0.05$ dex for stars in which the lithium doublet was detected. Our mean of all solar spectra is $\log\epsilon_{\rm Li}=1.03\pm0.04$ dex.

Initially, we derived Li abundances assuming line formation in LTE (local thermal equilibrium), in 1D, hydrostatic, plane parallel ATLAS 9 model atmospheres. Abundance corrections were thereafter applied to account for departures from LTE in the formation of the Li resonance line. The non-LTE modeling procedure is the same as described in \cite{Lind-Asplund-Barklem-09a}, but extended to cover also super-solar metallicities. For our sample stars, the abundance corrections range from -0.03 dex to +0.08 dex, depending on the lithium line strength and exact stellar parameters. In stars for which the equivalent width of the lithium line is below $\sim 100$ m$\AA$, over-ionization of neutral lithium results in positive abundances corrections, increasing with increasing metallicity and decreasing effective temperature. When the line starts to become saturated, increased photon losses pushes the statistical equilibrium in the opposite direction, i.e. into over-recombination, resulting in negative corrections for some stars \cite[see][for more details]{Lind-Asplund-Barklem-09a}. The non-LTE corrections are very small in comparison to the full range covered in lithium abundance, and hence the qualitative results of this study are the same for lithium abundances inferred in LTE and non-LTE. Note that the NLTE corrections are computed using MARCS models \cite[][]{Gustafsson-08}. Our NLTE corrected solar lithium abundance is $1.07\pm0.04$, in good agreement with the 3D-NLTE estimate by \cite{Asplund-09}.

\begin{figure}[ht]
   \centering
   \includegraphics[width=8.5cm]{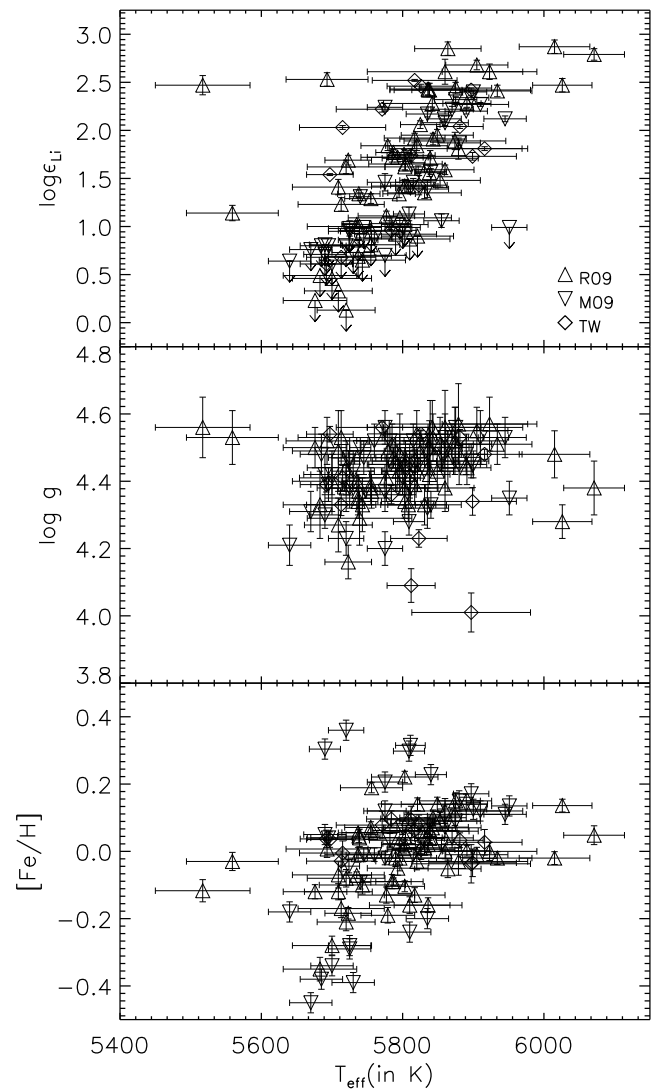}
      \caption{NLTE lithium abundance, $\log g$, and metallicity plotted against
      effective temperature. R09 and M09 stand for data from \cite{Ramirez-Melendez-Asplund-09}and \cite{Melendez-09, Melendez-10}, respectively; TW is data re-analyzed in this work.
              }
         \label{li-all}
   \end{figure}

Our adopted stellar parameters and derived LTE and non-LTE lithium abundances are given in Table \ref{table:parameters}. Fig. \ref{li-all} shows the good agreement between the three observational sub-samples, which reduces errors introduced by inhomogeneous data processing and therefore makes the analysis more reliable. It also is a proof of the consistently good quality of the data.

We have also considered the lithium abundances of solar twins from 8 open clusters as listed in Table~\ref{table:clusters}. Data are from the compilation by \cite{Sestito-Randich-05} as shown in Table~\ref{table:clusters} with updated data for M67 by \cite{Pasquini-08}. The age for M67 is taken from \cite{VandenBerg-07} and \cite{Yadav-08}, the lithium abundances for M67 stars are from \cite{Castro-10}. The clusters IC2602 and IC2391 are listed as one here, because their parameters are basically the same. We only used open clusters around solar metallicity ($0.0 \pm 0.1$ dex) and with reliable data. The solar twins that we select in open clusters are stars of one solar mass by definition, i.e. they are main sequence stars with 1M${_{\odot}}$ based on their effective temperature. We take into account the increase of the solar effective temperature with increasing age in the selection of stars from open clusters.

   \begin{figure}[ht]
   \centering
   \includegraphics[width=8.5cm]{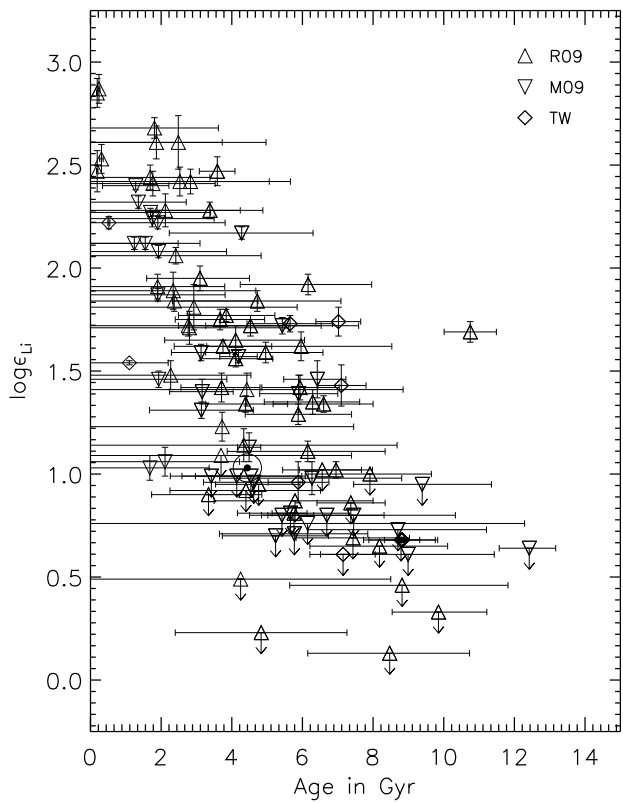}
      \caption{$\log\epsilon_{\rm Li}$ vs. age for stars from the three observational sub-samples.
      Down-arrows denote upper limits. The Sun is marked with $\odot$.
              }
         \label{age-li-all}
   \end{figure}

   \begin{figure}[ht]
   \centering
   \includegraphics[width=8.5cm]{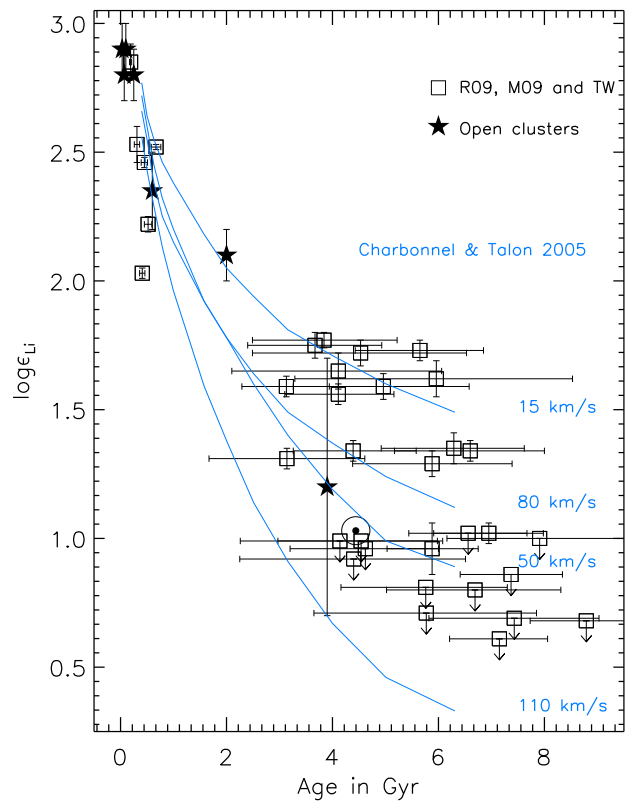}
      \caption{$\log\epsilon_{\rm Li}$ vs. age for solar twins from R09, M09, TW and from the solar twins in solar metallicity open clusters. Note the different scale compared to Fig. \ref{age-li-all}.
      The solid lines are the predicted values from the models by \cite{Charbonnel-Talon-05} for different initial rotational velocities.}
         \label{solar-twins}
   \end{figure}


\section{Results}\label{results}


\subsection{Lithium and age}

Using our sample of solar-like stars a clear lithium-age trend is observed (Fig. \ref{age-li-all}).
The dependency is as expected: older stars show more lithium depletion. The Spearman correlation coefficient is $r_{\rm tot}=-0.61$. This trend becomes clearer when we restrict the sample to solar twins, as in Fig. \ref{solar-twins}. We define solar twins as stars with [Fe/H] $=0.0\pm0.1$ and M $=(1.00\pm0.04)\mathrm{M}_{\odot}$ The stars from the open clusters given in Table \ref{table:clusters} fit the trend in Fig. \ref{solar-twins} very well. This is very important, because the ages of these clusters are well determined and the fact that they lie in the midst of the lithium vs. age trend of the field solar twins suggests that the ages we derived for individual stars are reliable. The Spearman correlation coefficient for the solar twin $\log \epsilon_{\rm Li}$-age trend including the open cluster data is $r_{\rm twin}=-0.75$. Another interesting thing to point out here is the fact that the Sun (marked with $\odot$ in the figures) fits the trend very well. This leads to the conclusion that the Sun does not have a particularly low lithium abundance compared to stars of similar age, mass, and metallicity.

Fig. \ref{solar-twins} also compares our observational data with model predictions from \cite{Charbonnel-Talon-05} for different initial rotational velocities of the stars. These hydrodynamical models give predictions for the evolution of surface Li abundance in solar-type stars, accounting self-consistently for element segregation and transport of angular momentum by rotation, gravity waves, and meridional circulation. They agree not only with the general lithium depletion trend observed by us, but it could also explain the relatively large scatter as a result of differences in initial stellar rotational velocities.

\begin{figure}[ht]
\centering
\includegraphics[width=8.5cm]{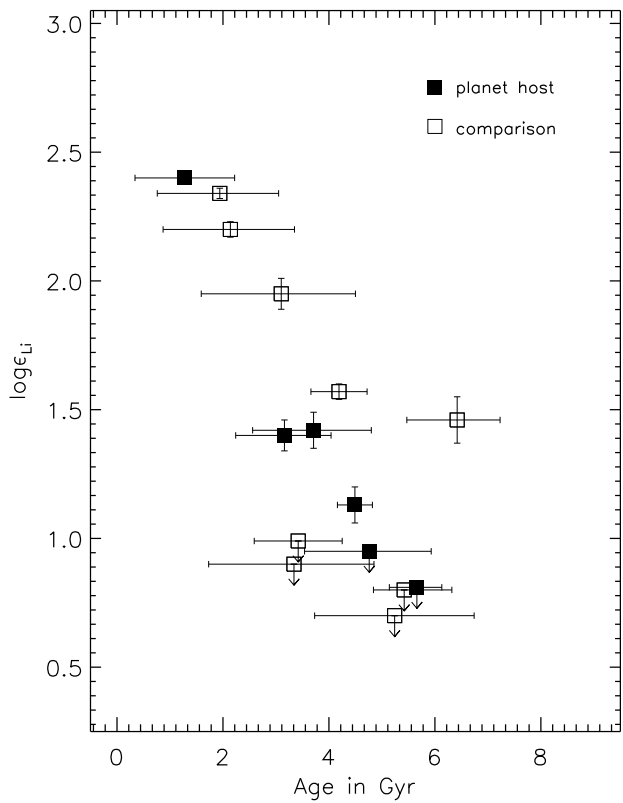}
   \caption{Same as Fig. \ref{solar-twins} but for metal-rich solar analogs ([Fe/H]$=0.25\pm0.15$, M$=1.08\pm0.08\,$M$_{\odot}$).}
      \label{solar-analogs}
\end{figure}


\subsection{Lithium and planets}

In Fig. \ref{solar-analogs} we plot lithium abundance against age, this time for a sample of metal-rich solar analogs. As metal-rich solar analogs we define stars with $[\mathrm{Fe/H}]=0.25\pm0.15$ and M $=(1.08\pm0.08)\mathrm{M}_{\odot}$. We use those criteria because our sub-sample of planet-hosts is biased towards those higher metallicities and masses. In this case we make a distinction between stars that are known to host planets (filled symbols) and those for which planets have not yet been detected (open symbols).

We used a two-dimensional Kolmogorov-Smirnov (KS) test to measure the probability that the samples of metal-rich solar analogs with and without planets in Fig. \ref{solar-analogs} belong to the same parent population. Using a Monte Carlo simulation, we took into account the errors in lithium abundance and age by choosing random, normally distributed values within each variable's 1-$\sigma$ environment on the linear scale, which means that instead of $\log\epsilon_{\rm Li}$, we used $10^{\log\epsilon_{\rm Li}-12}$, that is $\frac{n_{\rm Li}}{n_{\rm H}}$. The upper limits were accounted for by distributing the values uniformly between 0 and the upper limit.

We averaged the outcome of $1,000$ KS tests. This gave a probability for our metal-rich solar analogs with planets and those without planets to be part of the same parent sample of $64\pm15\%$; if we ignore the error bars and upper limits, this probability goes up to $80\%$. This is very important for the further analysis of the data, because it tells us that there is no \emph{intrinsic} difference between the two sub-samples. It is highly unlikely that the planet-hosts and comparison stars are different regarding their surface lithium abundance.

\begin{figure}[ht]
\centering
\includegraphics[width=8.5cm]{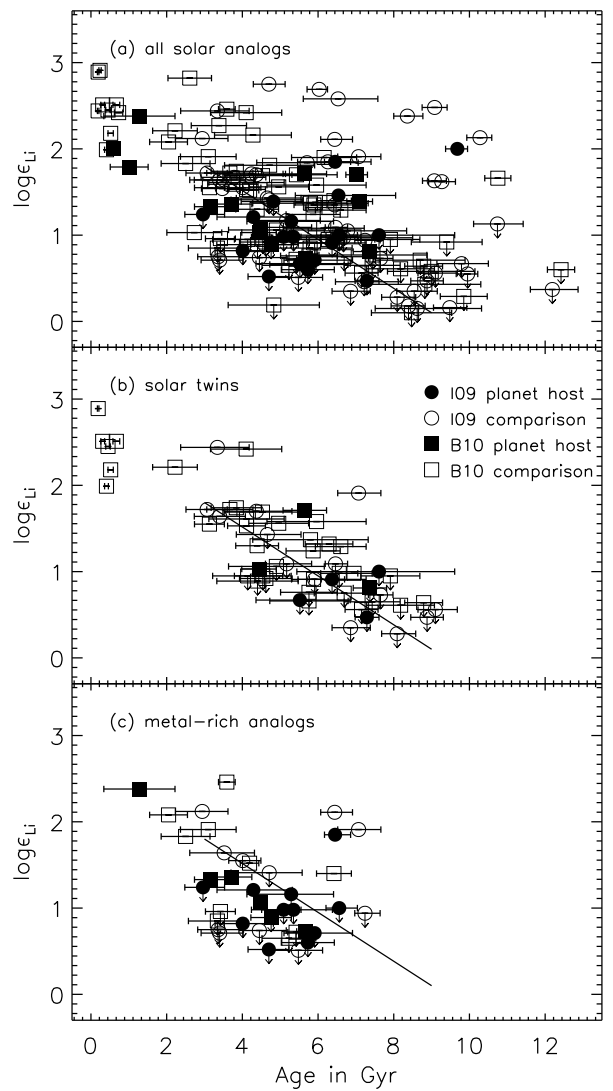}
   \caption{Comparison between the sample from \citet[][I09]{Israelian-09} and our sample (B10). The solid line, identically drawn in each panel, is an arbitrary reference line to guide the eye to the different $\log\epsilon_{\rm Li}$ levels in solar twins an metal-rich solar analogs. Note that for consistency we use the LTE $\log\epsilon_{\rm Li}$ values here.}
      \label{israelian}
\end{figure}

The age-lithium correlation coefficient for the solar twins is similar to that corresponding to the metal-rich solar analogs ($r_{\rm twin}=-0.75$, $r_{\rm analog}=-0.71$). However, the shapes of those trends are not identical. In the range from 3 to 6 Gyr, in particular, it is clear that for a given age, metal-rich solar analogs have on average lower lithium abundances than solar twins (see also Fig.~\ref{israelian} (c)). This is independent of whether the star has a planet or not. The age-lithium trend in Sun-like stars is thus metallicity dependent. This metallicity effect is predicted by stellar models due to the deeper convection zone in more metal-rich stars \cite[][]{Castro-09} and has lately been confirmed \cite[see, e.g.][Fig. 5]{doNascimento-daCosta-DeMedeiros-10}. Note, however, that the mass ranges being compared are somewhat different, and that this will have an impact on the lithium evolution of those two samples.

\begin{figure}[ht]
\centering
\includegraphics[width=8.5cm]{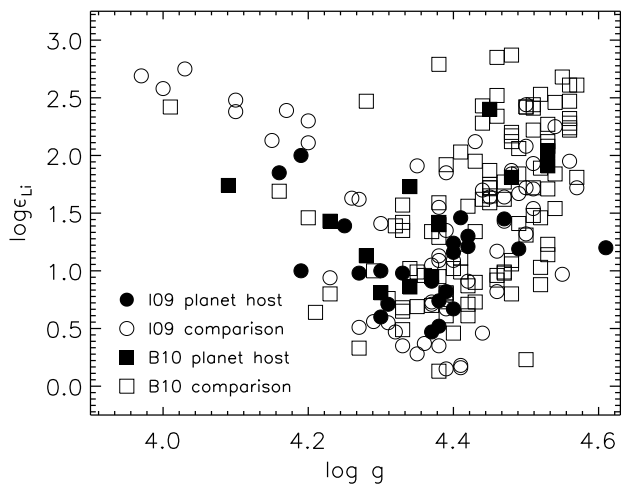}
   \caption{$\log\epsilon_{\rm Li}$ vs. $\log g$ for the stars from our sample (B10) and I09.}
   \label{logg}
\end{figure}


\section{Discussion}
Recently, it was claimed that planet-harboring solar-type stars show an enhanced lithium depletion and that lithium surface abundances in Sun-like stars do not correlate with stellar ages \cite[][hereafter I09 and S10, respectively]{Israelian-09, Sousa-10}.

In Fig. \ref{israelian}, we plot age versus lithium abundance for all 82 stars used in I09 along with the objects from this work (hereafter B10\footnote{For consistency, we used our LTE lithium abundance in this discussion because the I09 work does not take into account non-LTE corrections.}). With the stellar parameters adopted by I09 we derived the ages for that sample using the same techniques as for our sample; the ages we derive for the I09 sample are given in Table \ref{table:israelian}. Fig. \ref{israelian} shows the results for all stars panel (a), the solar twins (panel (b)), and the metal-rich solar analogs (panel (c)). We are using the same selection criteria for solar twins and metal-rich solar analogs as in Sect. \ref{results}. The agreement between the age-lithium relation found with our sample and that by I09 is excellent, in particular when looking at the solar twin plot. This shows that the stellar parameters used by I09 \cite[which were derived by][]{Sousa-08} are essentially on the same scale as ours and therefore the combination of both samples for this analysis does not introduce systematic errors. In fact, for the 10 stars in common between our sample and I09 we find differences of $3\pm20\,$K in T$_{\rm eff}$, $0.02\pm0.04$ in $\log g$, $0.003\pm0.023$ in [Fe/H], and $0.06\pm0.11$ in $\log\epsilon_{\rm Li}$ (the latter for the only 3 stars with lithium doublet detection).

In Fig. \ref{israelian}(a), ten stars with ages greater than 4 Gyr and higher than average $\log\epsilon_{\rm Li}$ can be seen above the main locus. Taking a closer look at those ``outliers'', the most interesting fact is that they have a particularly low surface gravity ($\log g \simeq 4.1$) compared to the rest of stars. In Fig. \ref{logg}, we show $\log\epsilon_{\rm Li}$ vs $\log g$. The surface lithium abundance on the low-$\log g$ side does not follow the main track, for which $\log\epsilon_{\rm Li}$ decreases with lower surface gravity, which is essentially the age effect, given that all these stars have similar masses. To exclude the possibility of systematic errors in the analysis of the low $\log g$ outliers, we compared the parameters and in particular their ages with various sources (see Table \ref{table:israelian_ages}). Our derived ages for these outliers are in reasonably good agreement with respect to the the values given in the literature. Only two stars appear to be older than the ages given in the consulted references, but even that difference is only around 2 Gyr and therefore not big enough for these stars to cease being outliers. This leads us to the conclusion that the ages we derived for the I09 sample and the stellar parameters adopted by I09 \cite[mostly derived by][]{Sousa-08} are correct and the high-lithium envelope in the lithium-age plane is most likely real. Thus, when restricted to a narrow range of T$_{\rm eff}$ around the solar value, $\log g\simeq4.1$ stars have higher lithium abundances than less evolved stars of similar age.\\

\begin{figure}[ht]
\centering
\includegraphics[width=8.5cm]{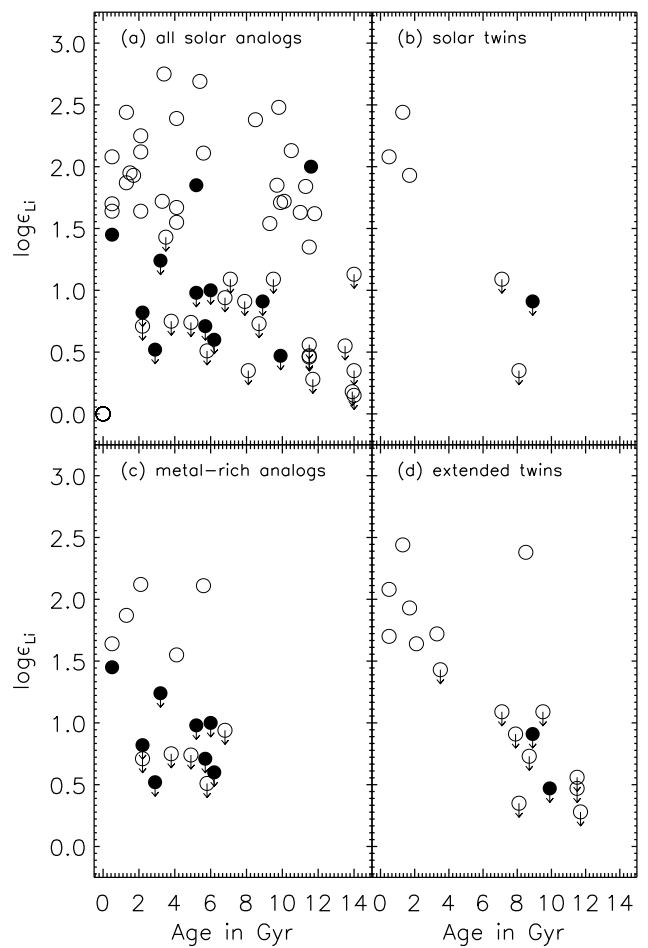}
   \caption{$\log\epsilon_{\rm Li}$ vs. age for the S10 sample. The selection criteria for the four panels are given in the text. Masses and ages adopted to make this figure are from S10.}
   \label{sousa}
\end{figure}

We have examined the results by S10, who claim that there is no correlation between lithium and age in the I09 sample. The S10 sample is basically the same as in I09, but reduced to the 60 stars studied in \cite{Sousa-08} because of the high homogeneity of the stellar parameters. We show their lithium-age trends in Fig. \ref{sousa}. The selection criteria we used for the solar twins and the metal-rich solar analogs are the same as in Fig. \ref{israelian}. However, this time we are using the masses and ages determined by S10. Although their full sample appears to show no correlation (Fig. \ref{sousa}a), the solar twin sample shows a clear trend between lithium and age. Note that the one planet-host in this sample has a low lithium abundance because of its old age and not the fact that it hosts a planet. There is no clear correlation between lithium and age for the metal-rich solar analog sample in this case, but this could be due to uncertain ages. Since the solar twin sample includes only 6 stars, we define another sample of ``extended solar twins'' with $[\mathrm{Fe/H}]=0.0\pm0.1$ and $M=1.00\pm0.10M_{\odot}$. The resulting figure shows a very definite trend of $\log\epsilon_{\rm Li}$ with age and only a single outlier appears. This outlier (HD215456), however, shows a relatively low $\log g$ of 4.10 (and an almost solar mass of 1.04M$_{\odot}$).\\

\begin{figure}[ht]
\centering
\includegraphics[width=8.5cm]{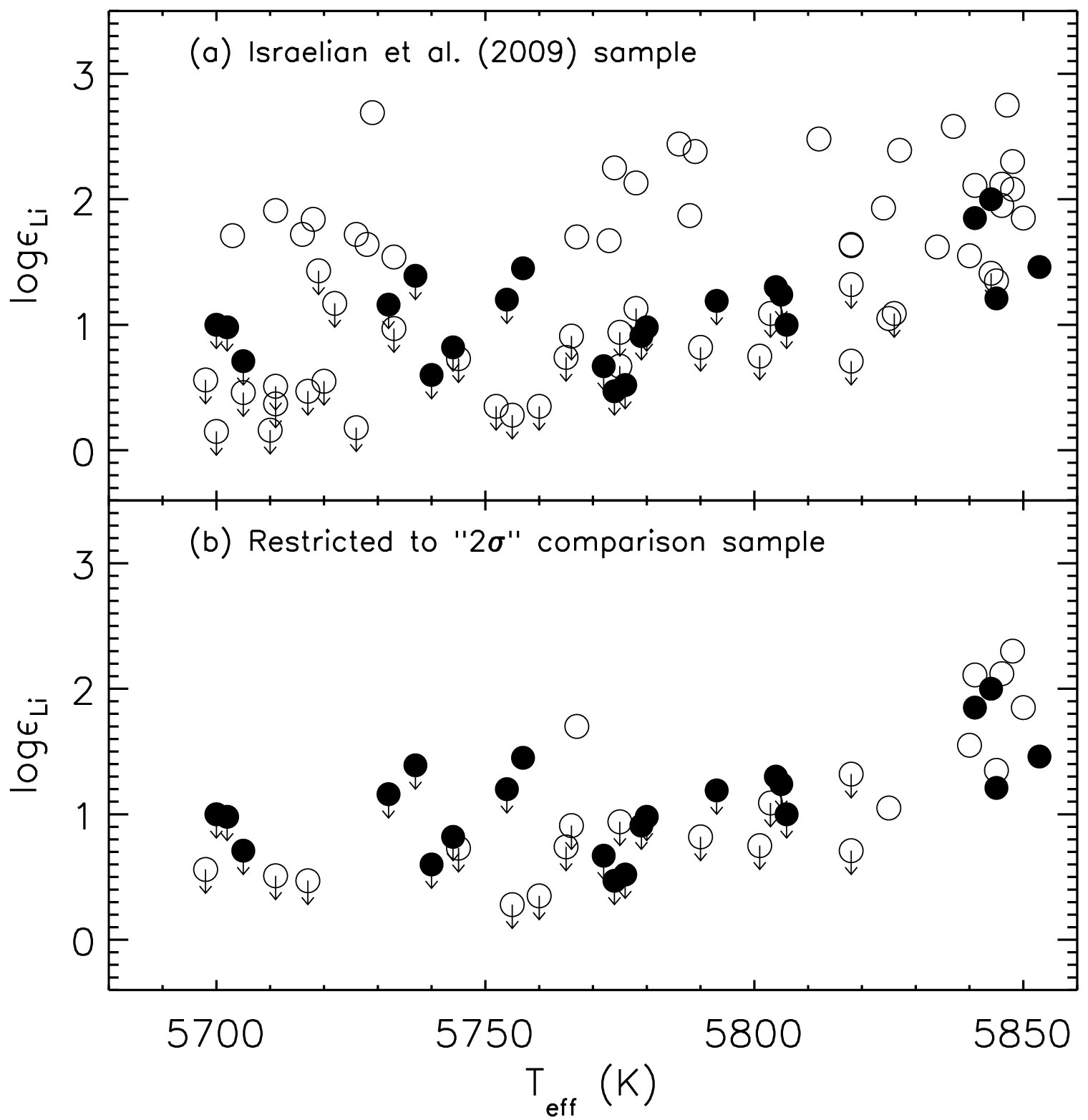}
   \caption{Lithium abundance as a function of $T_{\rm eff}$ in stars with and without detected planets from the I09 sample. Filled and open circles represent stars with and without detected planets, respectively. In the upper panel the original comparison made is shown, which is not appropriate because the sample being compared span different ranges in evolutionary phases and metallicities. A less biased comparison is shown in the bottom panel, where we only plot stars without detected planets with stellar parameters ($T_{\rm eff}, \log g$, [Fe/H]) within $2 \sigma$ of the planet-hosting stars. When a proper comparison is made, i.e., using stars with similar fundamental parameters, lithium is not abnormally low in stars with detected giant planets.}
   \label{teff-ali_is}
\end{figure}

We have also examined the lithium vs. effective temperature diagram presented by I09. As shown in Fig. \ref{teff-ali_is} (a), they found that almost all stars with a high lithium abundance ($\log\epsilon_{\rm Li}\gtrsim$1.5 dex) have not been shown to be planet hosts yet, whereas planet hosts and objects where no planets have been found are distributed quite equally below that lithium abundance, although the high number of upper limits makes a direct comparison difficult. In order to make a more robust comparison, we have restricted the comparison sample using the following criteria: we only considered comparison objects within a $2\sigma$ range in [Fe/H], log$g$, and T$_{\rm eff}$ around planet hosts, where $\sigma$ are the average values of the uncertainties in the stellar parameters given by \cite{Sousa-08}. In this way, we make sure that all stars lie within the same region of parameter space and are not influenced by the age or metallicity effects we find. Note that we do not restrict the lithium range, only metallicity, surface gravity and effective temperature. Using this selection allows for a homogeneous and unbiased comparison. When we restrict the parameter range covered by the comparison stars as described above, the lithium-planet connection disappears; as seen in Fig. \ref{teff-ali_is} (b), it is not possible to conclude on stronger lithium depletion in planet hosts compared to other stars. We stress that this figure is plotted directly from the I09 data without further manipulation or use of new parameters.

Three systematic biases have led I09 and S10 to conclude that solar-type planet-hosts feature an enhanced lithium depletion and that there is no age dependence:
\begin{enumerate}
\item At [Fe/H] $\simeq 0.0$, the existing HARPS sample of solar analogs with planets are on average older and therefore more depleted in lithium than non-planet-hosts, but not because they have planets.
\item At higher [Fe/H], where most of the I09 planet-hosts concentrate, there is a slightly different $\log\epsilon_{\rm Li}$ vs. age trend such that, at a given age in the $3-6\,$Gyr range, metal-rich solar analogs are more lithium-poor compared to solar metallicity ones. This is true for both planet-hosts as well as single stars.
\item I09 and S10 samples include a number of peculiarly high lithium abundance and relatively low $\log g\,(\simeq 4.1)$ stars; only one of them is a planet host.
\end{enumerate}
The apparently lower lithium abundances of planet-hosts found by I09 can thus be fully explained by a combination of age and metallicity effects, not separately but through the age vs. lithium relation.

\begin{table*}[!ht]
\begin{minipage}{170mm}
\begin{center}
\caption{Ages, masses, and lithium abundances for the outliers in Fig. \ref{israelian}(a).
	$R'_{HK}$ denotes ages derived from chromospheric activity, rot denotes ages derived from rotation
	periods.}
\label{table:israelian_ages}
\begin{tabular}{c c c c c}     
\hline\hline                 
object & age in Gyr & mass in M$_{\odot}$ & $\log\epsilon_{\rm Li}$ in dex & ages from other sources in Gyr$^a$\\
\hline
HD221420 & 4.70 & 1.30 & 2.75 & 4.5 (GCS), 4.1 (VF05),           \\
       &      &      &      & 5.1 (RP98)                         \\
HD114613 & 6.03 & 1.19 & 2.69 & 5.1 (RP98), 5.6 (B07,rot),       \\
       &      &      &      & 4.9 (VF05), 4.9 (RP98)             \\
HD2151   & 6.53 & 1.12 & 2.58 & 5.2 (GCS), 5.8 (VF05), 6.7 (V05) \\
HD215456 & 8.36 & 1.04 & 2.38 & 7.3 (GCS), 7.0 (I02)             \\
HD32724  & 9.07 & 0.97 & 1.63 & 9.9 (GCS)		         \\
HD4307   & 9.08 & 1.01 & 2.48 & 7.8 (W04, $R'_{HK}$, rot),       \\
       &      &      &      & 7.4 (GCS), 6.4 (VF05)              \\
HD78612  & 9.27 & 0.96 & 1.62 & 8.8 (GCS)		         \\
HD114729 & 9.68 & 0.97 & 2.00 & 10.9 (GCS), 6.45 (RP98)          \\
       &      &      &      & [planet-host]                      \\
HD145809 &10.28 & 0.96 & 2.13 & 6.9 (W04, $R'_{HK}$, rot),       \\
       &      &      &      & 7.9 (GCS), 7.4 (VF05)              \\
HD32923  &10.75 & 0.96 & 1.66 & 9.0 (VF05), 6.2 (W04),           \\
       &      &      &      & 9.9 (GCS), $>9.5$ (S83)            \\
\hline
\end{tabular}
\end{center}
$\!^a$The abbreviations used here are the following:\\
GCS: The Geneva-Copenhagen survey, \cite{Nordstrom-04}, VF05:
\cite{Valenti-Fischer-05}, RP98: \cite{Rocha-Pinto-Maciel-98}, V05: \cite{Vardavas-05}, I02:
\cite{Ibukiyama-Arimoto-02}, W04: \cite{Wright-04} and S83: \cite{Soderblom-83}).
\end{minipage}
\end{table*}

%

\section{Conclusions}

   \begin{enumerate}
      \item In stars of solar mass and solar metallicity, it is clear that older stars have experienced more surface lithium depletion. Both the overall lithium-age trend as well as the scatter that we observe in this sample of stars can be explained by the theoretical models by \cite{Charbonnel-Talon-05}.
      \item Metal-rich ([Fe/H]$\sim 0.25$) solar analogs (M$\sim 1.08$M$_{\odot}$) also exhibit a lithium-age trend, which is different from that seen in 1M$_{\odot}$, [Fe/H]$=0.0$ stars. At any given age in the 3 to 6 Gyr range, the metal-rich solar analogs are more lithium-poor. This is true for both planet-hosts and single stars.
      \item For solar-like stars, the lithium vs age trends for planet-hosts and stars where no planets have been found are statistically identical. Thus, the presence of a planet does not influence the observed surface lithium abundance.
      \item A number of solar-like stars with unusually high lithium abundance for their age are present in the field. We note that all of them have relatively low $\log g\simeq4.1$. We intend to pursue further observational work to better understand this small group of relatively low surface gravity and peculiarly high lithium abundance.
   \end{enumerate}

\begin{acknowledgements}
      We thank G. Israelian for sending us the lithium abundance data from I09.
\end{acknowledgements}

\bibliography{15137}
\bibliographystyle{aa}

\begin{sidewaystable*}\tiny
\begin{minipage}{170mm}
\begin{center}
\caption{Specifications for the different samples$^b$.}
\begin{tabular}{c c c c c c c c c c c c c c c c c c}     
\hline\hline                 
HIP & HD & $\mathrm{T}_{\mathrm{eff}}$ & $\sigma(\mathrm{T}_{\mathrm{eff}})$ & $\log g$ & $\sigma(\log g)$ & [Fe/H] &
$\sigma(\mathrm{[Fe/H]})$ & $\log\epsilon_{\rm Li}$ (LTE)& $\log\epsilon_{\rm Li}$ (NLTE)&
$\sigma(\log\epsilon_{\rm Li})$ & Mass & $\sigma$(m) & Age & $\sigma(\tau)$ & Source($\tau$) & Parameters & Planets\\
\hline
Sun   & Sun    & 5777 &   - & 4.44 &  -    &  0.000 &   -   & 1.03 & 1.07 &  0.04 & 1.00 & 0.01 &  4.5 & 0.5 & iso & -   & yes\\
348   & 225194 & 5777 &  40 & 4.41 & 0.07  & -0.130 & 0.024 & 1.08 & 1.11 &  0.05 & 0.97 & 0.01 &  6.2 & 1.5 & iso & R09 & no\\
996   & 804    & 5860 &  41 & 4.38 & 0.07  &  0.000 & 0.022 & 1.56 & 1.59 &  0.05 & 1.03 & 0.01 &  5.1 & 1.1 & iso & R09 & no\\
1499  & 1461   & 5756 &  44 & 4.37 & 0.05  &  0.189 & 0.015 & 0.89 & 0.95 & -1.00 & 1.06 & 0.01 &  4.8 & 0.6 & iso & R09+V05+LH06+T07+S08 & yes\\
2131  & 236416 & 5720 &  41 & 4.38 & 0.07  & -0.210 & 0.026 & 0.10 & 0.13 & -1.00 & 0.92 & 0.01 &  8.9 & 1.3 & iso & R09 & no\\
2894  & -      & 5820 &  44 & 4.54 & 0.07  & -0.030 & 0.025 & 1.81 & 1.84 &  0.05 & 1.02 & 0.01 &  1.6 & 0.8 & iso & R09 & no\\
4909  & 6204   & 5836 &  54 & 4.44 & 0.07  &  0.020 & 0.024 & 2.42 & 2.43 &  0.06 & 1.03 & 0.01 &  0.7 & 0.2 & x-ray & R09+G09 & no\\
5134  & 6470   & 5779 &  38 & 4.49 & 0.07  & -0.190 & 0.023 & 1.81 & 1.84 &  0.05 & 0.97 & 0.01 &  4.1 & 1.9 & iso & R09 & no\\
6407  & 8291   & 5787 &  25 & 4.47 & 0.03  & -0.090 & 0.011 & 1.74 & 1.77 &  0.03 & 0.99 & 0.01 &  3.8 & 0.8 & iso & R09 & no\\
7245  & 9446   & 5843 &  47 & 4.53 & 0.07  &  0.100 & 0.023 & 1.87 & 1.91 &  0.06 & 1.07 & 0.01 &  1.3 & 0.7 & iso & R09 & yes\\
8507  & 11195  & 5720 &  55 & 4.44 & 0.08  & -0.080 & 0.026 & 1.58 & 1.62 &  0.07 & 0.96 & 0.01 &  5.6 & 2.0 & iso & R09 & no\\
8841  & -      & 5676 &  45 & 4.50 & 0.06  & -0.120 & 0.021 & 0.19 & 0.23 & -1.00 & 0.95 & 0.01 &  4.3 & 1.8 & iso & R09 & no\\
9349  & 12264  & 5825 &  28 & 4.49 & 0.06  &  0.010 & 0.017 & 2.03 & 2.06 &  0.04 & 1.03 & 0.01 &  2.5 & 1.2 & iso & R09+T07 & no\\
10710 & -      & 5817 &  43 & 4.39 & 0.06  & -0.130 & 0.022 & 1.90 & 1.92 &  0.05 & 0.98 & 0.01 &  6.5 & 1.1 & iso & R09 & no\\
11072 & 14802  & 5897 &  84 & 4.01 & 0.058 & -0.037 & 0.057 & 2.42 & 2.42 &  0.02 & 1.15 & 0.02 &  0.7 & 0.1 & rot & TW+B07 & no\\
11728 & 15632  & 5738 &  30 & 4.37 & 0.05  &  0.045 & 0.019 & 1.29 & 1.34 &  0.04 & 0.99 & 0.01 &  6.9 & 0.8 & iso & R09+T07 & no\\
11915 & 16008  & 5793 &  43 & 4.45 & 0.06  & -0.050 & 0.021 & 1.69 & 1.72 &  0.05 & 1.00 & 0.01 &  4.3 & 1.6 & iso & R09 & no\\
12186 & 16417  & 5812 &  34 & 4.09 & 0.05  &  0.094 & 0.04  & 1.70 & 1.74 &  0.07 & 1.11 & 0.01 &  7.0 & 0.4 & iso & TW+V05+S08 & yes\\
14614 & 19518  & 5803 &  28 & 4.47 & 0.03  & -0.104 & 0.016 & 1.59 & 1.62 &  0.03 & 0.99 & 0.01 &  3.6 & 0.9 & iso & R09+T07+TW & no\\
14632 & 19373  & 6026 &  42 & 4.28 & 0.05  &  0.136 & 0.019 & 2.46 & 2.47 &  0.07 & 1.17 & 0.01 &  3.7 & 0.3 & iso & R09+V05 & unknown\\
15457 & 20630  & 5771 &  65 & 4.56 & 0.016 &  0.078 & 0.041 & 2.18 & 2.22 &  0.03 & 1.02 & 0.01 &  0.5 & 0.0 & rot & TW+V05 & unknown\\
18261 & 24552  & 5891 &  34 & 4.44 & 0.05  &  0.002 & 0.016 & 2.27 & 2.28 &  0.04 & 1.05 & 0.01 &  3.1 & 1.1 & iso & R09+T07 & no\\
22263 & 30495  & 5826 &  48 & 4.54 & 0.012 &  0.005 & 0.029 & 2.45 & 2.46 &  0.02 & 1.02 & 0.01 &  0.5 & 0.1 & rot & TW+V05 & unknown\\
22528 & -      & 5683 &  52 & 4.33 & 0.10  & -0.350 & 0.035 & 0.46 & 0.49 & -1.00 & 0.87 & 0.01 & 12.2 & 1.5 & iso & R09 & no\\
23835 & 32923  & 5723 &  33 & 4.16 & 0.05  & -0.184 & 0.017 & 1.66 & 1.69 &  0.05 & 0.96 & 0.01 & 10.7 & 0.4 & iso & R09+V05 & unknown\\
25670 & 36152  & 5755 &  37 & 4.38 & 0.05  &  0.071 & 0.017 & 1.24 & 1.29 &  0.05 & 1.01 & 0.01 &  6.0 & 0.9 & iso & R09+T07 & unknown\\
28336 & 40620  & 5713 &  61 & 4.53 & 0.08  & -0.170 & 0.027 & 1.19 & 1.23 &  0.07 & 0.95 & 0.01 &  4.0 & 1.8 & iso & R09 & no\\
29525 & 42807  & 5715 &  61 & 4.41 & 0.037 & -0.005 & 0.036 & 1.99 & 2.03 &  0.02 & 0.97 & 0.01 &  0.4 & 0.0 & rot & TW+B07 & no\\
30037 & 45021  & 5690 &  30 & 4.42 & 0.06  &  0.05  & 0.03  & 0.66 & 0.71 & -1.00 & 0.98 & 0.01 &  5.9 & 1.4 & iso & M09 & no\\
30502 & 45346  & 5745 &  25 & 4.47 & 0.05  & -0.01  & 0.02  & 0.95 & 0.99 & -1.00 & 1.00 & 0.01 &  3.8 & 1.4 & iso & M09 & no\\
36512 & 59711  & 5740 &  15 & 4.50 & 0.03  & -0.092 & 0.02  & 1.27 & 1.31 &  0.04 & 0.99 & 0.01 &  2.7 & 1.0 & iso & M09+T07+S08 & unknown\\
38072 & 63487  & 5839 &  68 & 4.53 & 0.11  &  0.060 & 0.037 & 1.67 & 1.71 &  0.08 & 1.05 & 0.01 &  2.4 & 1.2 & iso & R09 & no\\
38228 & 63433  & 5693 &  58 & 4.52 & 0.07  &  0.007 & 0.025 & 2.51 & 2.53 &  0.07 & 0.99 & 0.01 &  0.3 & 0.0 & rot & R09+V05, $\tau$ from G00& unknown\\
39748 & 67578  & 5835 &  30 & 4.48 & 0.06  & -0.20  & 0.03  & 2.16 & 2.17 &  0.03 & 0.98 & 0.01 &  3.8 & 1.6 & iso & M09 & no\\
41317 & 71334  & 5724 &  15 & 4.46 & 0.03  & -0.044 & 0.02  & 0.92 & 0.96 & -1.00 & 0.98 & 0.01 &  4.7 & 0.8 & iso & M09+V05+S08 & unknown\\
42438 & 72905  & 5864 &  47 & 4.46 & 0.09  & -0.052 & 0.026 & 2.89 & 2.85 &  0.07 & 1.02 & 0.01 &  0.2 & 0.0 & rot & R09+TW & no\\
43190 & 75288  & 5775 &  30 & 4.37 & 0.06  &  0.12  & 0.03  & 0.65 & 0.70 & -1.00 & 1.04 & 0.01 &  5.3 & 0.9 & iso & M09 & no\\
44324 & 77006  & 5934 &  49 & 4.51 & 0.06  & -0.020 & 0.019 & 2.41 & 2.41 &  0.06 & 1.07 & 0.01 &  1.5 & 0.6 & iso & R09+T07& no\\
44713 & 78429  & 5784 &  35 & 4.36 & 0.027 &  0.096 & 0.024 & 0.91 & 0.96 &  0.10 & 1.03 & 0.01 &  5.8 & 0.6 & iso & TW + VF05 + S08 & unknown\\
44935 & 78534  & 5800 &  25 & 4.41 & 0.05  &  0.07  & 0.02  & 0.95 & 0.99 & -1.00 & 1.03 & 0.01 &  4.6 & 1.0 & iso & M09 & no\\
44997 & 78660  & 5782 &  29 & 4.52 & 0.04  &  0.033 & 0.02  & 0.99 & 1.03 &  0.06 & 1.03 & 0.01 &  1.5 & 0.5 & iso & M09 + T07 & no\\
46066 & 80533  & 5709 &  65 & 4.49 & 0.12  & -0.070 & 0.039 & 1.37 & 1.41 &  0.08 & 0.96 & 0.01 &  5.2 & 2.3 & iso & R09 & no\\
46126 & 81700  & 5890 &  30 & 4.48 & 0.06  &  0.14  & 0.03  & 2.17 & 2.20 &  0.03 & 1.09 & 0.01 &  1.7 & 0.8 & iso & M09 & no\\
47990 & 84705  & 5910 &  40 & 4.53 & 0.08  &  0.12  & 0.03  & 2.24 & 2.27 &  0.02 & 1.10 & 0.01 &  0.8 & 0.4 & iso & M09 & no\\
49572 & -      & 5831 &  52 & 4.33 & 0.06  &  0.010 & 0.021 & 1.32 & 1.35 &  0.06 & 1.02 & 0.01 &  6.6 & 0.8 & iso & R09 & no\\
49756 & 88072  & 5804 &  52 & 4.45 & 0.07  &  0.041 & 0.023 & 1.61 & 1.65 &  0.07 & 1.03 & 0.01 &  3.6 & 1.6 & iso & R09+V05+T07 & unknown\\
50826 & -      & 5725 &  30 & 4.47 & 0.06  & -0.28  & 0.03  & 0.95 & 0.98 &  0.08 & 0.92 & 0.01 &  6.1 & 1.8 & iso & M09 & no\\
51258 & 90722  & 5720 &  25 & 4.23 & 0.05  &  0.360 & 0.03  & 0.72 & 0.80 & -1.00 & 1.17 & 0.02 &  5.1 & 0.2 & iso & M09+V05& unknown\\
52040 & 91909  & 5785 &  44 & 4.51 & 0.06  & -0.090 & 0.021 & 1.69 & 1.72 &  0.05 & 0.99 & 0.01 &  3.0 & 1.2 & iso & R09 & no\\
52137 & 92074  & 5842 &  69 & 4.56 & 0.08  &  0.070 & 0.026 & 2.25 & 2.28 &  0.08 & 1.06 & 0.01 &  0.9 & 0.5 & iso & R09 & no\\
53721 & 95128  & 5916 &  53 & 4.48 & 0.015 &  0.027 & 0.038 & 1.79 & 1.81 &  0.02 & 1.07 & 0.01 &  1.0 & 0.5 & iso + rot &  TW+V05+S05+M08 & yes\\
54102 & 96116  & 5870 &  30 & 4.51 & 0.06  &  0.04  & 0.03  & 2.20 & 2.22 &  0.03 & 1.06 & 0.01 &  1.6 & 0.7 & iso & M09 & no\\
55409 & 98649  & 5760 &  25 & 4.52 & 0.05  & -0.01  & 0.02  & 0.84 & 0.88 &  0.07 & 1.01 & 0.01 &  2.1 & 0.8 & iso & M09 & no\\
55459 & 98618  & 5838 &  21 & 4.42 & 0.03  &  0.038 & 0.012 & 1.53 & 1.56 &  0.04 & 1.03 & 0.01 &  4.1 & 0.7 & iso & R09+V05+M06+T07 & unknown\\
56948 & 101364 & 5795 &  23 & 4.43 & 0.03  &  0.023 & 0.014 & 1.30 & 1.34 &  0.04 & 1.01 & 0.01 &  4.4 & 0.7 & iso & R09+MR07+T09 & unknown\\
56997 & 101501 & 5559 &  65 & 4.53 & 0.08  & -0.030 & 0.027 & 1.08 & 1.14 &  0.08 & 0.94 & 0.01 &  4.6 & 2.1 & iso & R09+V05 & unknown\\
57291 & 102117 & 5690 &  22 & 4.30 & 0.04  &  0.304 & 0.03  & 0.73 & 0.81 & -1.00 & 1.11 & 0.01 &  5.6 & 0.2 & iso & M09+V05+S08& yes\\
59357 & 105779 & 5810 &  30 & 4.45 & 0.06  & -0.24  & 0.03  & 1.70 & 1.72 &  0.04 & 0.95 & 0.01 &  5.5 & 1.5 & iso & M09 & no\\
59610 & 106252 & 5899 &  62 & 4.34 & 0.041 & -0.034 & 0.041 & 1.71 & 1.73 &  0.04 & 1.04 & 0.01 &  5.7 & 0.7 & iso & TW+V05 & yes\\
60081 & 107148 & 5811 &  21 & 4.38 & 0.04  &  0.315 & 0.03  & 1.33 & 1.40 &  0.06 & 1.12 & 0.01 &  3.4 & 0.5 & iso & M09+V05+S08 & yes\\
60314 & 107633 & 5874 &  72 & 4.52 & 0.10  &  0.110 & 0.033 & 1.85 & 1.89 &  0.09 & 1.07 & 0.01 &  1.9 & 0.9 & iso & R09 & no\\
\hline
\end{tabular}
\end{center}
\end{minipage}									       
\end{sidewaystable*}

\begin{sidewaystable*}\tiny
\begin{minipage}{170mm}
\begin{center}
\begin{tabular}{c c c c c c c c c c c c c c c c c c}
\hline\hline                 
HIP & HD & $\mathrm{T}_{\mathrm{eff}}$ & $\sigma(\mathrm{T}_{\mathrm{eff}})$ & $\log g$ & $\sigma(\log g)$ & [Fe/H] &
$\sigma(\mathrm{[Fe/H]})$ & $\log\epsilon_{\rm Li}$ (LTE)& $\log\epsilon_{\rm Li}$ (NLTE)&
$\sigma(\log\epsilon_{\rm Li})$ & Mass & $\sigma$(m) & Age & $\sigma(\tau)$ & Source($\tau$) & Parameters & Planets\\
\hline
60370  & 107692 & 5897 &  25 & 4.46 & 0.05  &  0.171 & 0.03  & 2.31 & 2.34 &  0.02 & 1.11 & 0.01 &  1.6 & 0.8 & iso & M09+V05 & unknown\\
60653  & 108204 & 5725 &  30 & 4.38 & 0.06  & -0.29  & 0.03  & 0.92 & 0.95 & -1.00 & 0.90 & 0.01 &  9.6 & 1.2 & iso & M09 & no\\
62175  & 110869 & 5849 &  51 & 4.43 & 0.06  &  0.140 & 0.021 & 1.91 & 1.95 &  0.06 & 1.08 & 0.01 &  2.9 & 1.2 & iso & R09+T07 & no\\
64150  & 114174 & 5755 &  41 & 4.39 & 0.05  &  0.056 & 0.016 & 0.76 & 0.81 & -1.00 & 1.00 & 0.01 &  5.9 & 1.0 & iso & R09+V05+T07 & unknown\\
64497  & 114826 & 5860 & 110 & 4.56 & 0.11  &  0.120 & 0.037 & 2.60 & 2.61 &  0.13 & 1.07 & 0.02 &  1.3 & 0.8 & iso & R09 & no\\
64713  & 115169 & 5815 &  25 & 4.52 & 0.05  & -0.01  & 0.02  & 1.43 & 1.46 &  0.04 & 1.03 & 0.01 &  1.7 & 0.8 & iso & M09 & no\\
64794  & 115382 & 5743 &  61 & 4.33 & 0.08  & -0.100 & 0.027 & 0.61 & 0.65 & -1.00 & 0.96 & 0.01 &  8.8 & 1.0 & iso & R09 & no\\
64993  & 115739 & 5875 &  30 & 4.56 & 0.06  &  0.09  & 0.03  & 2.30 & 2.32 &  0.03 & 1.08 & 0.01 &  0.9 & 0.5 & iso & M09 & no\\
66618  & 118475 & 5951 &  25 & 4.35 & 0.05  &  0.135 & 0.03  & 0.96 & 0.99 & -1.00 & 1.12 & 0.01 &  3.6 & 0.5 & iso & M09+V05 & unknown\\
66885  & 119205 & 5685 &  30 & 4.48 & 0.06  & -0.38  & 0.03  & 0.77 & 0.80 & -1.00 & 0.88 & 0.01 &  7.4 & 1.9 & iso & M09 & no\\
69063  & 123152 & 5670 &  30 & 4.31 & 0.06  & -0.45  & 0.03  & 0.74 & 0.76 & -1.00 & 0.83 & 0.01 & 14.1 & 0.7 & iso & M09 & no\\
71683  & 128620 & 5840 &  22 & 4.33 & 0.04  &  0.228 & 0.03  & 1.52 & 1.57 &  0.03 & 1.11 & 0.01 &  4.3 & 0.3 & iso & M09+V05+PM08& unknown\\
72659  & 131156 & 5517 &  67 & 4.56 & 0.09  & -0.117 & 0.033 & 2.44 & 2.47 &  0.10 & 0.91 & 0.01 &  0.2 & 0.0 & rot & R09+V05 & unknown\\
73815  & 133600 & 5803 &  33 & 4.34 & 0.05  &  0.020 & 0.016 & 0.98 & 1.02 & -1.00 & 1.01 & 0.01 &  6.8 & 0.6 & iso & R09+MR07 & no\\
74341  & 134902 & 5853 &  57 & 4.51 & 0.08  &  0.090 & 0.026 & 1.44 & 1.48 &  0.07 & 1.07 & 0.01 &  1.9 & 0.9 & iso & R09 & no\\
74389  & 134664 & 5859 &  24 & 4.48 & 0.04  &  0.105 & 0.03  & 2.08 & 2.12 &  0.03 & 1.07 & 0.01 &  1.6 & 0.7 & iso & M09+S08 & unknown\\
75923  & 138159 & 5775 &  25 & 4.56 & 0.05  & -0.02  & 0.02  & 2.21 & 2.24 &  0.04 & 1.02 & 0.01 &  0.9 & 0.2 & iso & M09 & no\\
77052  & 140538 & 5697 &  33 & 4.54 & 0.023 &  0.035 & 0.023 & 1.49 & 1.54 &  0.01 & 1.01 & 0.01 &  0.6 & 0.4 & iso & TW+V05 & unknown\\
77466  & -	& 5700 &  56 & 4.40 & 0.09  & -0.280 & 0.028 & 0.43 & 0.46 & -1.00 & 0.90 & 0.01 &  9.3 & 1.9 & iso & R09 & no\\
77740  & 141937 & 5900 &  19 & 4.45 & 0.04  &  0.125 & 0.03  & 2.38 & 2.40 &  0.02 & 1.09 & 0.01 &  1.3 & 0.9 & iso + rot & M09+M05+S05+V05+LH06+S08 & yes\\
77883  & 142331 & 5695 &  25 & 4.39 & 0.05  &  0.04  & 0.02  & 0.75 & 0.80 & -1.00 & 0.98 & 0.01 &  7.0 & 0.9 & iso & M09 & no\\
78028  & -	& 5879 &  98 & 4.57 & 0.12  & -0.030 & 0.041 & 1.79 & 1.81 &  0.11 & 1.04 & 0.02 &  1.8 & 1.0 & iso & R09 & no\\
78680  & 144270 & 5923 &  67 & 4.57 & 0.08  & -0.000 & 0.027 & 2.61 & 2.61 &  0.08 & 1.06 & 0.01 &  0.6 & 0.4 & iso & R09 & no\\
79186  & 145514 & 5709 &  48 & 4.27 & 0.08  & -0.120 & 0.024 & 0.29 & 0.33 & -1.00 & 0.95 & 0.01 & 10.3 & 0.7 & iso & R09 & no\\
79304  & 145478 & 5945 &  30 & 4.53 & 0.06  &  0.11  & 0.03  & 2.09 & 2.12 &  0.03 & 1.11 & 0.01 &  0.4 & 0.2 & iso & M09 & no\\
79578  & 145825 & 5860 &  33 & 4.53 & 0.07  &  0.072 & 0.03  & 2.05 & 2.08 &  0.03 & 1.07 & 0.01 &  1.3 & 0.6 & iso & M09+V05 & unknown\\
79672  & 146233 & 5822 &   9 & 4.45 & 0.02  &  0.051 & 0.02  & 1.55 & 1.59 &  0.04 & 1.04 & 0.01 &  3.1 & 0.5 & iso & M09+TW+S08& unknown\\
80337  & 147513 & 5881 &  33 & 4.53 & 0.024 &  0.033 & 0.022 & 2.01 & 2.04 &  0.02 & 1.06 & 0.03 &  0.6 & 0.1 & rot & TW  (S$^4$N+HARPS)+V05+S08+B07 & yes\\
81512  & -	& 5790 &  58 & 4.46 & 0.07  & -0.020 & 0.025 & 0.89 & 0.92 & -1.00 & 1.01 & 0.01 &  4.0 & 1.7 & iso & R09 & no\\
82853  & 150027 & 5640 &  30 & 4.21 & 0.06  & -0.18  & 0.03  & 0.60 & 0.64 & -1.00 & 0.92 & 0.01 & 12.5 & 0.5 & iso & M09 & no\\
83601  & 154417 & 6071 &  43 & 4.38 & 0.08  &  0.048 & 0.028 & 2.82 & 2.79 &  0.06 & 1.13 & 0.01 &  2.4 & 1.1 & iso & R09+V05 & unknown\\
83707  & 152441 & 5880 &  30 & 4.45 & 0.06  &  0.15  & 0.03  & 1.83 & 1.87 &  0.03 & 1.10 & 0.01 &  2.3 & 1.0 & iso & M09 & no\\
85042  & 157347 & 5692 &  37 & 4.39 & 0.022 &  0.037 & 0.026 & 0.56 & 0.61 & -1.00 & 0.98 & 0.01 &  7.2 & 0.5 & iso & TW+V05+S08 & unknown\\
85272  & 156922 & 5700 &  30 & 4.42 & 0.06  & -0.34  & 0.03  & 0.58 & 0.61 & -1.00 & 0.88 & 0.01 &  9.4 & 1.4 & iso & M09 & no\\
85285  & 157691 & 5730 &  30 & 4.43 & 0.06  & -0.39  & 0.03  & 0.71 & 0.73 & -1.00 & 0.88 & 0.01 &  8.8 & 1.5 & iso & M09 & no\\
86796  & 160691 & 5809 &  22 & 4.28 & 0.04  &  0.298 & 0.03  & 1.06 & 1.13 &  0.07 & 1.15 & 0.01 &  4.6 & 0.2 & iso & M09+V05+S08& yes\\
88194  & 164595 & 5735 &  21 & 4.40 & 0.03  & -0.071 & 0.010 & 0.98 & 1.02 &  0.04 & 0.96 & 0.01 &  7.3 & 0.6 & iso & R09+V05+T07 & unknown\\
88427  & -	& 5810 &  57 & 4.42 & 0.07  & -0.160 & 0.025 & 0.85 & 0.87 & -1.00 & 0.97 & 0.01 &  5.7 & 1.5 & iso & R09 & no\\
89162  & 165357 & 5835 &  30 & 4.32 & 0.06  &  0.07  & 0.03  & 1.35 & 1.39 &  0.09 & 1.05 & 0.01 &  6.0 & 0.6 & iso & M09 & no\\
89443  & 238838 & 5796 &  73 & 4.48 & 0.12  & -0.020 & 0.038 & 1.06 & 1.09 & -1.00 & 1.01 & 0.01 &  4.4 & 1.9 & iso & R09 & no\\
89650  & 167060 & 5855 &  25 & 4.48 & 0.05  &  0.02  & 0.02  & 1.03 & 1.06 &  0.07 & 1.05 & 0.01 &  2.2 & 1.0 & iso & M09 & no\\
91332  & 171918 & 5775 &  25 & 4.20 & 0.05  &  0.206 & 0.03  & 1.40 & 1.46 &  0.09 & 1.12 & 0.02 &  6.4 & 0.7 & iso & M09+V05 & unknown\\
96402  & 184768 & 5713 &  49 & 4.33 & 0.032 & -0.029 & 0.030 & 0.64 & 0.68 & -1.00 & 0.97 & 0.01 &  8.7 & 0.7 & iso & TW+T07 & no\\
96901  & 186427 & 5737 &  28 & 4.34 & 0.04  &  0.055 & 0.016 & 1.38 & 1.42 & -1.00 & 1.00 & 0.01 &  7.5 & 0.5 & iso & R09+V05+LH06+T07 & yes\\
96895  & 186408 & 5808 &  39 & 4.33 & 0.05  &  0.097 & 0.020 & 0.81 & 0.86 &  0.06 & 1.05 & 0.01 &  6.0 & 0.6 & iso & R09+V05+LH06 & unknown\\
100963 & 195034 & 5802 &  17 & 4.45 & 0.03  &  0.008 & 0.013 & 1.72 & 1.75 &  0.05 & 1.02 & 0.01 &  3.6 & 0.8 & iso & R09+T07+T09 & no\\
100970 & 195019 & 5823 &  40 & 4.23 & 0.026 &  0.083 & 0.025 & 1.39 & 1.43 &  0.10 & 1.06 & 0.01 &  7.2 & 0.4 & iso & TW+V05 & yes\\
102152 & 197027 & 5737 &  47 & 4.35 & 0.06  & -0.010 & 0.022 & 0.65 & 0.69 & -1.00 & 0.98 & 0.01 &  7.7 & 0.9 & iso & R09+M09 & no\\
104504 & 201422 & 5836 &  48 & 4.50 & 0.06  & -0.160 & 0.022 & 2.42 & 2.42 &  0.06 & 1.00 & 0.02 &  3.0 & 1.4 & iso & R09  & no\\
107350 & 206860 & 6015 &  50 & 4.48 & 0.07  & -0.020 & 0.019 & 2.91 & 2.87 &  0.07 & 1.09 & 0.01 &  0.2 & 0.0 & rot & R09+V05 & unknown\\
108708 & 209096 & 5875 &  51 & 4.51 & 0.07  &  0.150 & 0.024 & 2.42 & 2.44 &  0.06 & 1.10 & 0.01 &  1.3 & 0.6 & iso & R09 & no\\
108996 & 209562 & 5838 &  56 & 4.50 & 0.08  &  0.060 & 0.027 & 2.40 & 2.42 &  0.07 & 1.05 & 0.01 &  2.3 & 1.1 & iso & R09 & no\\
109110 & 209779 & 5817 &  60 & 4.46 & 0.033 &  0.062 & 0.030 & 2.51 & 2.52 &  0.01 & 1.04 & 0.01 &  0.7 & 0.1 & rot & TW+V05+T07+B07 & unknown\\
109931 & -      & 5739 &  74 & 4.29 & 0.08  &  0.040 & 0.026 & 0.95 & 1.00 & -1.00 & 1.00 & 0.01 &  8.2 & 0.9 & iso & R09 & no\\
113357 & 217014 & 5803 &  47 & 4.38 & 0.05  &  0.221 & 0.017 & 1.36 & 1.42 &  0.07 & 1.09 & 0.01 &  3.9 & 0.7 & iso & R09+V05+LH06 & yes\\
115604 & -      & 5821 &  51 & 4.43 & 0.06  &  0.140 & 0.019 & 0.85 & 0.90 & -1.00 & 1.07 & 0.01 &  3.1 & 1.2 & iso & R09 & no\\
118159 & 224448 & 5905 &  44 & 4.55 & 0.07  & -0.010 & 0.022 & 2.69 & 2.68 &  0.05 & 1.06 & 0.01 &  0.8 & 0.4 & iso & R09 & no\\    
\hline				  											       
\end{tabular} 													       
\label{table:parameters}
\end{center}
$\!^b$The abbreviations of the sources in Table \ref{table:parameters} are the following:\\
R09, M09 and TW are from \cite{Ramirez-Melendez-Asplund-09}, \cite{Melendez-09},
\cite{Melendez-10} and this work, as before. M06 is \cite{Melendez-Dodds-Eden-Robles-06},
V05 is \cite{Valenti-Fischer-05}, S08 is \cite{Sousa-08}, B07 \cite{Barnes-07},
LH06 \cite{Luck-Heiter-06}, PM08 \cite{PortodeMello-Lyra-Keller-08}, T07 and T09 are
\cite{Takeda-07} and \cite{Takeda-Tajitsu-09}, respectively, G00 is
\cite{Gaidos-Henry-Henry-00} and MR07 is \cite{Melendez-Ramirez-07}. S05 is
\cite{Saffe-Gomez-Chavero-05}, G09 is \cite{Guinan-Engle-09}, and M08 \cite{Mamajek-Hillenbrand-08}.\\
The $-1$ in $\sigma(\log\epsilon_{\rm Li})$ denotes upper limits.
\end{minipage}
\end{sidewaystable*}
\normalsize

\begin{table*}\tiny
\begin{center}
\caption{Sample used in I09. Masses and ages are from this work.}
\begin{tabular}{c c c c c c | c c c c c c c}     
\hline\hline                 
Star name or HIP & HD & Mass & $\sigma$(m) & Age & $\sigma(\tau)$ & &
Star name or HIP & HD & Mass & $\sigma$(m) & Age & $\sigma(\tau)$ \\
\hline
WASP 5 & -     & 0.99 & 0.06 &  7.9 & 3.3 &  & 52409  & 92788  & 1.08 & 0.01 &  3.8 &  1.0\\
XO-1   & -     & 1.01 & 0.01 &  2.3 & 1.2 &  & 53837  & 95521  & 0.98 & 0.01 &  3.4 &  1.1\\ 
1499  & 1461   & 1.07 & 0.01 &  4.5 & 0.6 &  & 54287  & 96423  & 1.01 & 0.01 &  7.2 &  0.6\\ 
1954  & 2071   & 0.97 & 0.01 &  4.6 & 1.1 &  & 54400  & 96700  & 0.97 & 0.01 &  6.6 &  0.6\\ 
2021  & 2151   & 1.12 & 0.08 &  6.7 & 1.4 &  & 97998  & 97998  & 0.90 & 0.01 &  1.9 &  0.6\\ 
5339  & 4307   & 1.01 & 0.01 &  9.0 & 0.4 &  & 60081  & 107148 & 1.12 & 0.01 &  3.0 &  0.6\\ 
6455  & 8406   & 0.98 & 0.01 &  3.0 & 0.6 &  & 60729  & 108309 & 1.05 & 0.01 &  7.5 &  0.3\\ 
8798  & 11505  & 0.93 & 0.01 &  8.6 & 0.6 &  & 62345  & 111031 & 1.10 & 0.01 &  3.4 &  0.8\\ 
9381  & 12387  & 0.91 & 0.01 &  9.1 & 1.7 &  & 64408  & 114613 & 1.20 & 0.02 &  5.9 &  0.3\\ 
9683  & 12661  & 1.10 & 0.03 &  4.5 & 1.3 &  & 64459  & 114729 & 0.97 & 0.01 &  9.7 &  0.2\\ 
12048 & 16141  & 1.09 & 0.01 &  6.9 & 0.3 &  & 64550  & 114853 & 0.92 & 0.01 &  7.4 &  0.9\\ 
12186 & 16417  & 1.12 & 0.01 &  6.7 & 0.2 &  & 65036  & 115585 & 1.13 & 0.03 &  5.3 &  0.5\\ 
14501 & 19467  & 0.94 & 0.01 & 10.0 & 0.3 &  & 71683  & 128620 & 1.17 & 0.07 &  4.1 &  1.5\\ 
15442 & 20619  & 0.94 & 0.01 &  3.9 & 1.2 &  & 74500  & 134987 & 1.10 & 0.02 &  5.4 &  0.5\\ 
15330 & 20766  & 0.94 & 0.02 &  3.4 & 1.7 &  & 78330  & 143114 & 0.88 & 0.01 &  9.9 &  0.8\\ 
15527 & 20782  & 0.98 & 0.01 &  7.3 & 0.3 &  & 78459  & 143761 & 0.98 & 0.02 &  6.1 &  2.6\\ 
16365 & 21938  & 0.86 & 0.01 & 10.8 & 0.7 &  & 79524  & 145809 & 0.96 & 0.01 & 10.3 &  0.3\\ 
19925 & 27063  & 1.01 & 0.01 &  4.2 & 1.2 &  & 79672  & 146233 & 1.03 & 0.01 &  3.3 &  0.8\\ 
20625 & 28471  & 0.97 & 0.01 &  7.7 & 0.3 &  & 83906  & 154962 & 1.22 & 0.03 &  4.7 &  0.8\\ 
20677 & 28701  & 0.89 & 0.01 &  9.5 & 0.5 &  & 160691 & 160691 & 1.14 & 0.02 &  4.8 &  0.3\\ 
23627 & 32724  & 0.97 & 0.01 &  9.2 & 0.3 &  & 95962  & 183658 & 1.01 & 0.01 &  5.3 &  0.7\\ 
22504 & 34449  & 1.02 & 0.01 &  1.5 & 0.8 &  & 96901  & 186427 & 1.02 & 0.02 &  5.0 &  1.9\\ 
25670 & 36152  & 1.05 & 0.01 &  2.6 & 0.9 &  & 97336  & 187123 & 1.07 & 0.01 &  3.5 &  1.5\\ 
26737 & 37962  & 0.94 & 0.01 &  5.2 & 1.8 &  & 97769  & 188015 & 1.10 & 0.02 &  1.8 &  0.9\\ 
27435 & 38858  & 0.95 & 0.01 &  3.3 & 0.7 &  & 98959  & 189567 & 0.92 & 0.01 &  8.4 &  0.4\\ 
30243 & 44420  & 1.11 & 0.01 &  3.5 & 0.6 &  & 98589  & 189625 & 1.09 & 0.01 &  2.5 &  1.0\\ 
30104 & 44594  & 1.08 & 0.00 &  4.1 & 0.5 &  & 102664 & 198075 & 0.99 & 0.01 &  2.3 &  1.0\\ 
30476 & 45289  & 0.97 & 0.00 &  8.8 & 0.3 &  & 104903 & 202206 & 1.09 & 0.01 &  1.4 &  0.6\\ 
34065 & 53705  & 0.97 & 0.01 &  6.8 & 2.3 &  & 106006 & 204313 & 1.06 & 0.01 &  4.6 &  0.5\\ 
36512 & 59711A & 0.96 & 0.01 &  5.3 & 1.0 &  & 108468 & 208704 & 0.99 & 0.01 &  6.6 &  0.3\\ 
39417 & 66428  & 1.09 & 0.02 &  5.8 & 1.0 &  & 109821 & 210918 & 0.96 & 0.01 &  8.2 &  0.4\\ 
43726 & 76151  & 1.05 & 0.01 &  1.5 & 0.5 &  & 110109 & 211415 & 0.96 & 0.01 &  6.5 &  1.2\\ 
43686 & 76700  & 1.17 & 0.07 &  4.5 & 1.2 &  & 112414 & 215456 & 1.04 & 0.01 &  8.4 &  0.4\\ 
44713 & 78429  & 1.02 & 0.01 &  7.0 & 0.5 &  & 113357 & 217014 & 1.08 & 0.02 &  3.4 &  1.6\\ 
44890 & 78538  & 1.01 & 0.01 &  2.5 & 1.1 &  & -      & 219542 & 1.04 & 0.02 &  4.6 &  1.5\\ 
44860 & 78558  & 0.85 & 0.01 & 12.5 & 0.7 &  & 115577 & 220507 & 0.98 & 0.01 &  9.3 &  0.5\\ 
44896 & 78612  & 0.96 & 0.01 &  9.4 & 0.3 &  & 116250 & 221420 & 1.29 & 0.06 &  4.7 &  0.7\\ 
46007 & 81110  & 1.11 & 0.01 &  0.4 & 0.1 &  & 116852 & 222480 & 1.15 & 0.03 &  5.6 &  0.8\\ 
49728 & 88084  & 0.97 & 0.01 &  6.2 & 0.8 &  & 116906 & 222582 & 0.99 & 0.01 &  6.7 &  0.8\\ 
50534 & 89454  & 1.03 & 0.01 &  3.0 & 1.1 &  & 117320 & 223171 & 1.09 & 0.01 &  6.7 &  0.3\\ 
52369 & 92719  & 1.01 & 0.01 &  1.6 & 0.9 &  & 118123 & 224393 & 0.92 & 0.01 &  3.6 &  1.0\\
\hline												     
\end{tabular}											     
\label{table:israelian} 									     
\end{center}											     
\end{table*}

\end{document}